\documentclass[prl,aps,twocolumn,showpacs,floatfix]{revtex4-2}
\usepackage{amsfonts}
\usepackage{amssymb}
\usepackage{hyperref}
\usepackage{graphicx}
\usepackage{dcolumn}
\usepackage{bm,amsmath,verbatim}
\usepackage{mathrsfs}
\usepackage{color}
\usepackage{mathtools}

\usepackage{amsmath,amssymb,amsthm}

\usepackage[T1]{fontenc}
\usepackage[latin9]{inputenc}
\usepackage{booktabs}

\hypersetup{colorlinks,
	linkcolor=blue,          citecolor=blue,        filecolor=blue,      urlcolor=blue           }

\newcommand{\ket}[1]{|#1\rangle}
\newcommand{\bra}[1]{\langle #1 |}

\newcommand{\ii}{\text{i}}

\begin{document}

    \title{Dynamical Transition due to Feedback-induced Skin Effect}
    \author{Ze-Chuan Liu$^{1}$}
    \author{Kai Li$^{1}$}
    \author{Yong Xu$^{1,2}$}
    \email{yongxuphy@tsinghua.edu.cn}
    \affiliation{$^{1}$Center for Quantum Information, IIIS, Tsinghua University, Beijing 100084, People's Republic of China}
    \affiliation{$^{2}$Hefei National Laboratory, Hefei 230088, PR China}

\begin{abstract}
The traditional dynamical phase transition refers to the appearance of singularities in an observable with respect to
a control parameter for a late-time state or singularities in the rate function of the Loschmidt echo with respect to time.
Here, we study the many-body dynamics in a continuously monitored free fermion system with conditional
feedback under open boundary conditions.
We surprisingly find a novel dynamical transition
from a logarithmic scaling of the entanglement entropy to an area-law scaling as time evolves.
The transition, which is noticeably different from the conventional dynamical phase transition,
arises from the competition between the bulk dynamics and boundary skin effects.
In addition, we find that while quasidisorder or disorder cannot drive a transition for the steady state,
a transition occurs for the maximum entanglement entropy during the time evolution,
which agrees well with the entanglement transition for the steady state of the dynamics under periodic boundary conditions.
\end{abstract}
\maketitle

Non-Hermitian physics has sparked considerable interests from various fields in recent
years due to the presence of intriguing phenomena~\cite{ChristodoulidesNPReview,XuReview,UedaReview,BergholtzReview},
such as exceptional points or rings~\cite{Moiseyev2011,Zhen2015nat,Xu2017PRL,Nori2017PRL,Kawabata2019PRL}
and non-Hermitian skin effects~\cite{HN1996PRL,HN1997PRB,TonyLee,Yao2018PRL1,Xiong2018JPC,Torres2018PRB,Kunst2018PRL}. In fact,
non-Hermiticity is ubiquitous in the dynamics of many-body quantum systems~\cite{QMCbook}.
For instance, when considering postselection in continuous measurements, the many-body dynamics is dictated by a
non-Hermitian Hamiltonian~\cite{QMCbook}. Through non-Hermitian many-body dynamics, many interesting non-equilibrium
phenomena have been discovered~\cite{Ueda2018PRL, Ueda2019PRL, Ueda2020PRL, Lucas2020PRR, Balazs2021PRB, Gullans2021PRL, Chen2021PRB, Imura2022PRB, Marco2023Scipost, Ryu2022arXiv, Schiro2023PRB, Pal2023arxiv,Dreyer2023PRL,Longhi2023PRB,Imura2023arxiv, Larson2023arxiv}.
In particular, it has been found that the steady state of a non-Hermitian
dynamics of free fermions in the Hatano-Nelson (HN) model is the skin state where particles mainly reside
at a half of a system, possessing an area-law scaling of the entanglement entropy~\cite{Ryu2022arXiv}. Later,
it was shown that disorder or quasidisorder can drive an entanglement phase transition for the steady state with peculiar
scaling properties~\cite{Kai2023, ShanZhong2023,LongWen2023}.

However, to observe these interesting phenomena, one has to apply
postselection, which is formidably challenging for a large system. Fortunately, non-Hermitian dynamics arises
naturally in quantum dissipative systems or continuously monitored systems, as dictated by the Lindblad master equation,
due to the typically non-Hermitian nature of the Liouvillian that governs the system's dynamics.
Interestingly, the Liouvillian eigenvectors can also exhibit skin effects~\cite{Song2019PRL, Haga2021PRL, Yang2022PRR}.
In addition, Wang {\it et al.} studied the dynamics of a quantum system under continuous measurements
with conditional feedback~\cite{ChenFang2022arXiv} and found that the late-time steady state always concentrates at an
edge~\cite{ChenFang2022arXiv,ShuChen2022arXiv}. 

In the field of non-equilibrium many-body dynamics, one central question to ask is whether there exists
a dynamical phase transition. Such a transition is classified into two types.
One type corresponds to the appearance of a non-analytical behavior in an observable
with respect to a control parameter for a late-time state~\cite{Altshuler2006PRL,Kollar2008PRL,Kehrein2008PRL,Werner2009PRL,Biroli2010PRL,Rey2022RPP}.
Roughly speaking, measurement induced entanglement phase transitions~\cite{Fisher2018PRB, Smith2019PRB, Nahum2019PRX, Fisher2019PRB, Altman2020PRL, Diehl2021PRL, Schiro2021PRB, Schiro2021Quantum, Saito2022PRL, Luca2019Scipost,Yao2022PRL, Buchhold2022PRL,Xiaoliang2022, 
Turkeshi2022PRB, Buchhold2022PRR, Schiro2022PRB, Diehl2022arxiv, Passarelli2023arxiv, Hamazaki2023arxiv, Xing2024PRB} may fall into this category when
the entanglement entropy is regarded as an observable and the measurement rate as a control parameter.
The other type of dynamical phase transition occurs if the rate function of the return probability develops a
singularity as time evolves~\cite{Heyl2013PRL,Silva2018PRL,Heyl2018RPP,Zvyagin2016LTP,Fabrizio2016PTRA}.
Interestingly, it has been shown that the presence of skin effects precludes the existence of the
entanglement phase transition for the steady state~\cite{ChenFang2022arXiv,ShuChen2022arXiv}. We therefore ask whether
the many-body skin effects can lead to the second type dynamical phase transition or other novel dynamical phase transition
beyond this traditional one with respect to time.

\begin{figure}[t]
    \includegraphics[width=1\linewidth]{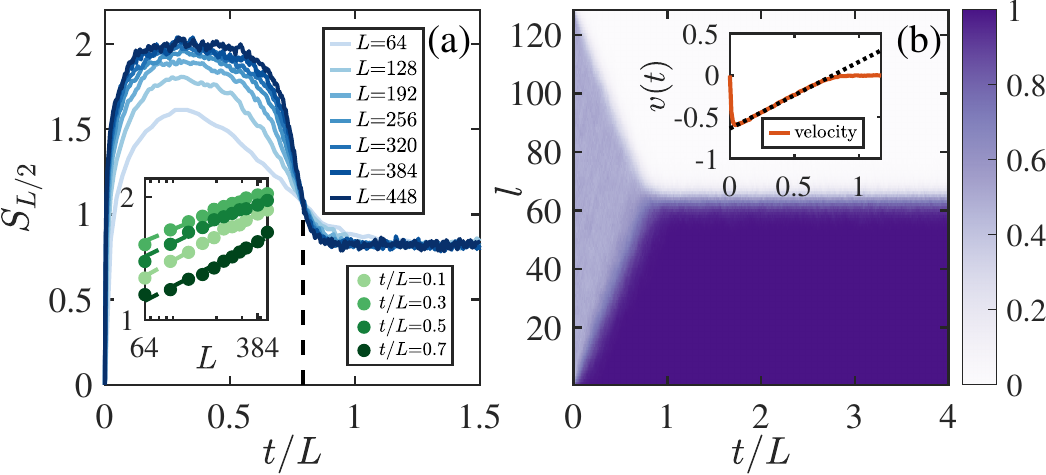}
    \caption{(a) The time evolution of the trajectory averaged bipartite entanglement entropy $S_{L/2}$ for different
        system sizes $L$ ($t$ denotes the evolution time). The black dashed line highlights the transition point
        between the logarithmic-law and area-law regimes.
        The inset shows the data for several $t/L$ values in a linear-log scale with $L$
        up to 448, illustrating the log-law scaling.
        (b) The time evolution of the trajectory averaged density distribution with respect to site $l$ for
        a system with $L=128$. Inset: the average velocity ${v}(t)$ under OBCs as a function
        of $t/L$ (red line). The black dashed line represents a linear fit of ${v}(t)$
        before the transition point. }
    \label{Fig1}
\end{figure}%%%

Here we study the many-body dynamics of a continuously monitored free fermion system
under open boundary conditions (OBCs).
Surprisingly, we find the existence of a novel dynamical transition from a logarithmic scaling
of the entanglement entropy to an area-law scaling with respect to $t/L$ where $t$ is the evolution time and
$L$ is the system size (see Fig.~\ref{Fig1}), although the late-time states approach the skin state even in the presence of
strong quasidisorder [see Fig.~\ref{Fig3}(d)]. The dynamical transition arises due to the competition between the bulk dynamics and
the dynamics significantly affected by edges. The competition can also be seen from the fact that this dynamical transition
does not happen under periodic boundary conditions (PBCs).
It is important to note that this transition is fundamentally different from the traditional
two types of dynamical phase transitions.
In our case, instead of the return probability, the probability for the evolving state to be found in an ideal many-body skin state
exhibits a sharp rise across the transition point.
In addition, we investigate whether quasidisorder can induce an entanglement transition under
OBCs in this system.
We find that for the late-time steady state, the transition does not happen.
However, we show that the dynamics at earlier times is mainly governed by the bulk,
resulting in a rapid growth of the entanglement entropy. Later, the entanglement entropy
reaches a maximum followed by a decline to a stable value due to the strong boundary effects caused by skin effects.
Remarkably, we find that these maximum entanglements undergo a transition from a logarithmic to an area-law
scaling as we raise the quasidisorder strength. The results agree well with the transition observed for the
steady states under PBCs.

\emph{Trajectory evolution}.---
We start by considering a spinless fermion chain of length $L$ with
nearest-neighbor hopping and an onsite quasiperiodic potential described by the Hamiltonian,
\begin{equation} \label{Eq:Hamiltonian}
    \hat{H}=\sum_{l}\left( J(\hat{c}_l^\dagger \hat{c}_{l+1}+\text{H.c.})
    +W\cos{(2\pi\alpha l )}\hat{c}_l^\dagger \hat{c}_{l} \right),
\end{equation}
where $\hat{c}_l^\dagger\ \left(\hat{c}_l\right)$ is the creation (annihilation) operator of a fermion at site $l$,
$J$ denotes the hopping strength (we set $J=1$ as units of energy hereafter), and $W$ represents the
amplitude of the potential, which is quasiperiodic when $\alpha$ takes irrational values
(we consider $\alpha=\left(\sqrt{5}-1\right)/2$ without loss of generality).
To generate skin effects, we follow Refs.~\cite{ChenFang2022arXiv,ShuChen2022arXiv} to introduce continuous measurements with feedback
represented by the quantum jump operator
$\hat{L}_l=e^{i\theta\hat{n}_{l+1}}\hat{d}_l^\dagger \hat{d}_l$ with the strength proportional
to $\gamma$, where $\hat{n}_l=\hat{c}_l^\dagger \hat{c}_l$,
$\hat{d}_l=\frac{1}{\sqrt{2}}(\hat{c}_l+i\hat{c}_{l+1})$, and $\theta$ is a parameter that
takes real values (we set $\theta=\pi$ hereafter).
The evolution trajectory of a pure state $|\psi_t\rangle$ at time $t$ is then described by a
stochastic Schr\"{o}dinger equation (SSE)~\cite{QMCbook,supplement},
\begin{equation} \label{Eq:SDE}
d|\psi_t\rangle=-i\hat{H}_{\mathrm{eff}}|\psi_t\rangle dt+\sum_l \left({\hat{L}_l}/\sqrt{a_0}-1\right)|\psi_t\rangle dW_l,
\end{equation}
where $\hat{H}_{\mathrm{eff}}=\hat{H}-i\frac{\gamma}{2}\sum_l \hat{L}^\dagger_l\hat{L}_l $ is
the effective Hamiltonian, $a_0=\langle \psi_t|\hat{L}^\dagger_l\hat{L}_l |\psi_t\rangle $,
and $\left\{dW_l\right\}$ denotes a set of Poisson random numbers
that take the values of either 0 or 1 with the mean value of $\overline{dW_l}=\gamma a_0 dt$.
Without loss of generality, we take $\gamma=0.5$ for all subsequent calculations.
Previous studies have shown that when $\theta \neq 0$ (e.g., $\theta=\pi$), the steady state is always
the skin state for a system under OBCs since the detected right-moving particles will be converted to the left-moving ones
through the imposed phase~\cite{ChenFang2022arXiv,ShuChen2022arXiv}.
In contrast to these studies, we will examine whether a transition appears as time evolves.

To study the dynamical transition, we initialize the system in the Ne\' {e}l state, that is,
$|\psi_0\rangle =\prod_{l=1}^{L/2}\hat{c}_{2l}^\dagger |0\rangle$, where $|0\rangle$ is the vacuum state.
We then numerically calculate each trajectory of $|\psi_t\rangle$ by solving Eq.~(\ref{Eq:SDE}).
The observables are calculated in each trajectory at every time slice, and then averaged over all trajectories.
Given that the effective Hamiltonian is quadratic and the jump operators have a simple form,
the dynamics could be efficiently simulated with a Slater determinant state~\cite{supplement}.
Furthermore, the Gaussian structure of the states allows for the extraction of observables from
the correlation matrix $C_{ij}(t)=\langle \psi_t|\hat{c}^\dagger_ic_j|\psi_t\rangle$. To capture the
entanglement transition, we focus on the von Neumann entanglement entropy calculated by~\cite{Peschel2003JPA},
%\begin{equation} \label{Eq:EE}
$S_A=-[\textrm{Tr}\left(C_A \log{C_A}+(1-C_A) \log{(1-C_A)}\right)]$,
%\end{equation}
where $C_{A,mn} =\langle \hat{c}_m^\dagger\hat{c}_n\rangle$ with $m,n \in A$
represents the correlation matrix of a subsystem $A$ of an evolving state.
Hereafter, we use $[\dots]$ to denote the average over
trajectories (our numerical results are averaged over 180-2000 trajectories unless stated otherwise)
and use $S_{L/2}$ to denote the
entanglement entropy of the left bipartition of the chain.

\begin{figure}
\includegraphics[width=1\linewidth]{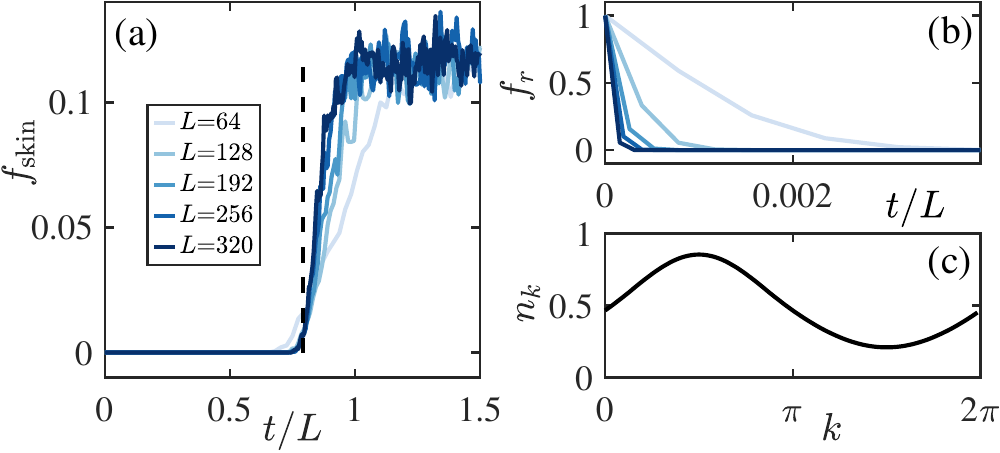}
\caption{The time evolution of (a) the probability $f_{\mathrm{skin}}$ of the evolving state $|\psi_t\rangle$
    being in an ideal many-body skin state $|\psi_{\mathrm{skin} }\rangle$ and (b) the return probability
    $f_r$ for different system sizes averaged over trajectories. In (b), we consider 96 trajectories.
    The black dashed line marks the transition point estimated in Fig.~\ref{Fig1}(a).
    (c) The density distribution with respect to the momentum $k$ for the steady
    state of the dynamics under PBCs with $L=128$.
}
\label{Fig2}
\end{figure}%%%

\emph{Dynamical transition}.---
We now investigate the
entanglement transition that occurs during dynamics for a system under OBCs
without quasidisorder (i.e., $W=0$).
To compare the dynamics of systems with different sizes $L$, we rescale the time $t$ to $t/L$.
Remarkably, we find that the trajectory averaged entanglement entropy $S_{L/2}$
undergoes a transition from a log-law
to an area-law regime at $t_c/L \approx 0.79$ as shown in Fig.~\ref{Fig1}(a).
Specifically, as time evolves, $S_{L/2}$ first increases, then decreases, and eventually
reaches a steady value, which is independent of system sizes. The log-law scaling before
the transition is shown in the inset of Fig.~\ref{Fig1}(a).

To explain the behavior of $S_{L/2}$, we examine the evolution of the density distribution
$[\langle \hat{n}_l\rangle]$. As shown in Fig.~\ref{Fig1}(b), all particles move to the left side
of the chain and eventually populate the left part. This phenomenon is a result of the feedback-induced
skin effect, which has been discussed in Refs.~\cite{ChenFang2022arXiv,ShuChen2022arXiv}.
However, the dynamical transition has not been found there.
By a comparison with
the evolution of $S_{L/2}$,
we observe that before all particles reach the left side, $S_{L/2}$ increases with increasing $L$.
Once the state $|\psi_t\rangle$ becomes a skin state, $S_{L/2}$ no longer grows.

\begin{figure}
\includegraphics[width=1\linewidth]{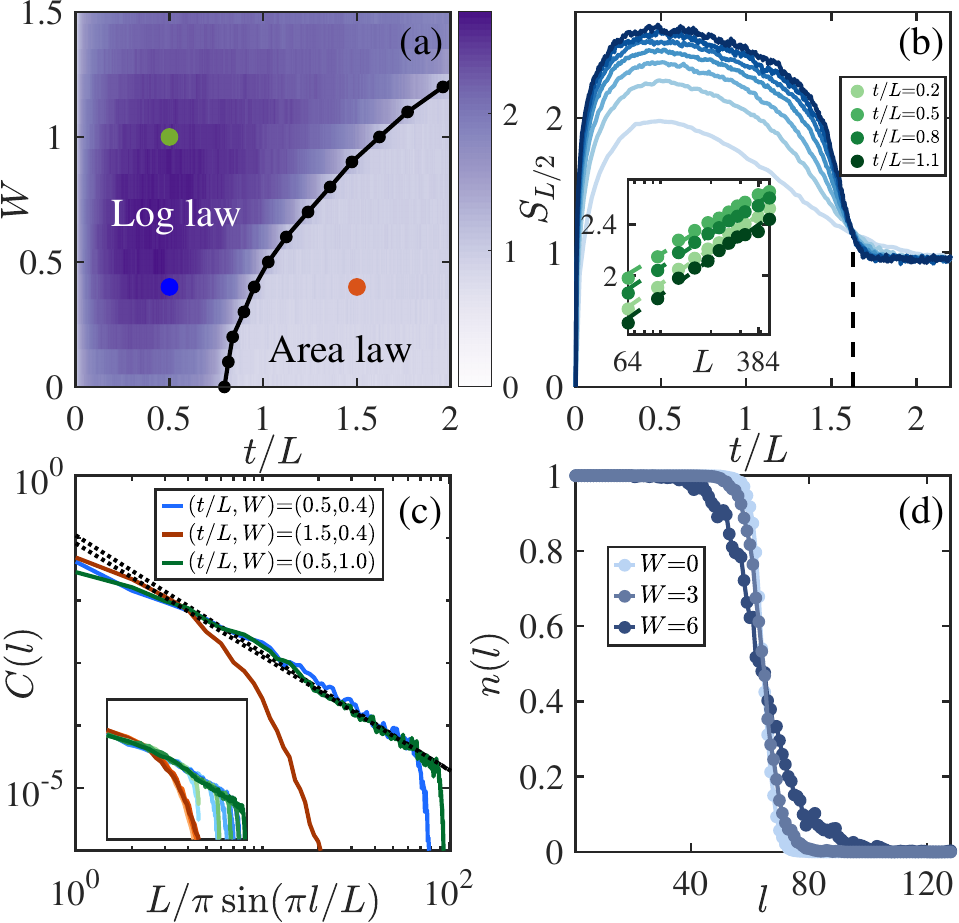}
\caption{(a) The phase diagram of the trajectory averaged bipartite entanglement entropy
    in the plane $(t/L,W)$ with $L=320$.
    The black line represents the dynamical transition line, separating the log-law regime and the area-law regime.
    (b) The trajectory averaged entanglement entropy $S_{L/2}$ versus $t/L$ at $W=1$
    with $L=64, 128, 192, 256, 320, 384, 448$ (from light to dark colors).
    Inset: the linear-log scale plot of $S_{L/2}$ versus $L$ at $t/L=0.2, 0.5, 0.8, 1.1$ with a log-law fit (the dashed lines).
    (c) The correlation functions $C(l)$ versus $(L/\pi)\sin{(\pi l/L)}$ at different $(t/L,W)$ marked by solid blue,
    red and green circles in (a), respectively. Here $L=320$.
    The inset displays a data collapse for different system sizes $L=64,128,192,256,320$, with the same
    axes range as the main plot.
    (d) The steady-state density distribution ${n}(l)$ at $W=0, 3, 6$. The total evolution time is set to 10000 to
    ensure that the final state reaches the steady state.
}
\label{Fig3}
\end{figure}%%%

We note that this dynamical transition is different from
the conventional dynamical phase transition that exhibits singularities in the rate function of
the Loschmidt echo at the transition point~\cite{Heyl2018RPP}. In our dynamics, the return probability
$f_r=[|\langle \psi_0 |\psi_t\rangle|^2]$ suddenly drops to zero at
a very short time as shown in Fig.~\ref{Fig2}(b). Interestingly, we find that the
probability $f_{\mathrm{skin}}=[ |\langle\psi_{\mathrm{skin}}|\psi_t\rangle|^2 ]$ of the evolving state being in an ideal many-body skin state
$|\psi_{\mathrm{skin}}\rangle=\prod_{l=1}^{L/2}\hat{c}_{l}^\dagger|0\rangle$
experiences a sudden rise at the transition point
[Fig.~\ref{Fig2}(a)], indicating that the skin state develops through a dynamical transition.

We expect that the dynamical transition arises from the competition between the dynamics under PBCs and
boundary skin effects. At earlier times, since most particles are far away from a boundary,
their time evolution is mainly governed by the system under PBCs. However, as time progresses, particles
approach a boundary so that the boundary effect becomes relevant.

To demonstrate the competition, we analyze the average velocity per particle
${v}(t)=d [\langle {\hat{x}}\rangle ]/{dt}$ under OBCs
where $[\langle{\hat{x}}\rangle]=2\sum_l l [\langle \psi_t|\hat{n}_l |\psi_t\rangle ] /L$ is the average position of
the particles~\cite{supplement}.
If we assume that the velocity of a single left-moving particle is $v_0$, then
a fraction of $2|v_0|t/L$ fermionic particles will reach the left side at time $t$ ($t<t_c$). As a result,
only the remaining particles contribute to the velocity. The average velocity before the transition is
thus given by
%\begin{equation} \label{Eq:vmanyevol}
${v}(t) = \left (1-2|v_0|t/L\right )v_0$.
%\end{equation}
We plot ${v}(t)$ in the inset of Fig.~\ref{Fig1}(b), where the black dashed line represents
a linear fit of the data. The slope of the line is determined to be $2v^2_0=0.801$, from which we
obtain $v_0 = -0.633$. Since the particles with the velocity $v_0$ are within the bulk, their properties
(including velocities) should be determined by the system under PBCs as we have expected.
We now use $v_0=({2}/{L})\sum_k  {n}_k  v_k$
to estimate $v_0$ through pure bulk dynamics.
Here, $v_k=-2\sin{k}$ is the group velocity of the free fermion Hamiltonian at the momentum $k$,
and ${n}_k=[\langle \psi_{\mathrm{steady}} | \hat{c}_k^\dagger \hat{c}_k |\psi_{\mathrm{steady}}\rangle ]$
is the density of particles at
$k$ in the steady state $|\psi_{\mathrm{steady}}\rangle$
of the dynamics under PBCs [see Fig.~\ref{Fig2}(c)].
The estimated velocity $v_0$ is $-0.633$, which is the same as the velocity calculated under OBCs.
The critical rescaled time $t_c/L$ is determined by
the fact that all particles move to the left side at this time so that $v(t_c)=0$. Thus, $t_c/L=1/(2|v_0|)=0.79$,
which is in excellent agreement with the results in Fig.~\ref{Fig1}(a).
The transition is reminiscent of the kinetic phase transition where the average velocity
vanishes with respect to a control parameter such as noise~\cite{Vicsek1995PRL,Vicsek1999PRL}. In contrast,
our transition occurs with respect to a rescaled time.

\emph{Dynamical transition with quasiperiodic potential}.---
We now investigate the dynamical entanglement transition that occurs in the presence of the quasiperiodic
potential.
By numerically analyzing the entanglement entropy $S_{L/2}$ at different times and
for various values of $L$ and $W$, we map out the phase diagram in Fig.~\ref{Fig3}(a).
We see that as $W$ increases, the dynamical
entanglement transition persists, with the transition point being delayed.
Specifically, we plot the evolution of $S_{L/2}$ for various values of $L$ when $W=1$
in Fig.~\ref{Fig3}(b), clearly illustrating that $S_{L/2}$ transitions from the logarithmic scaling
[see the inset of Fig.~\ref{Fig3}(b)] to the area-law scaling. At the transition point, $t_c/L\approx 1.63$, which is larger than the transition
point value in the case without quasidisorder. The delayed transition point is attributed to the fact
that quasidisorder slows down the particles' motion, but eventually,
the particles migrate to the left boundary. In addition, we find that the probability of the evolving state
being in an ideal many-body skin state
suddenly rises across the transition point~\cite{supplement}, similar to the case without disorder.

To further provide evidence for the dynamical transition, we calculate the
connected density-density correlation function
%\begin{equation}\label{Eq:cor}
$C(l)=[ \langle \hat{n}_{L/2}\rangle\langle \hat{n}_{L/2+l}\rangle-\langle \hat{n}_{L/2}\hat{n}_{L/2+l}\rangle ]$
%\end{equation}
for different values of $W$ and $t/L$ and plot them in Fig.~\ref{Fig3}(c).
We see clearly that $C(l)$ decays algebraically when the point $(W,t/L)$ falls within the log-law regime,
whereas it decays exponentially in the area-law regime. This change in the behavior of $C(l)$ further
reveals the existence of the dynamical transition.

\begin{figure}
\includegraphics[width=1\linewidth]{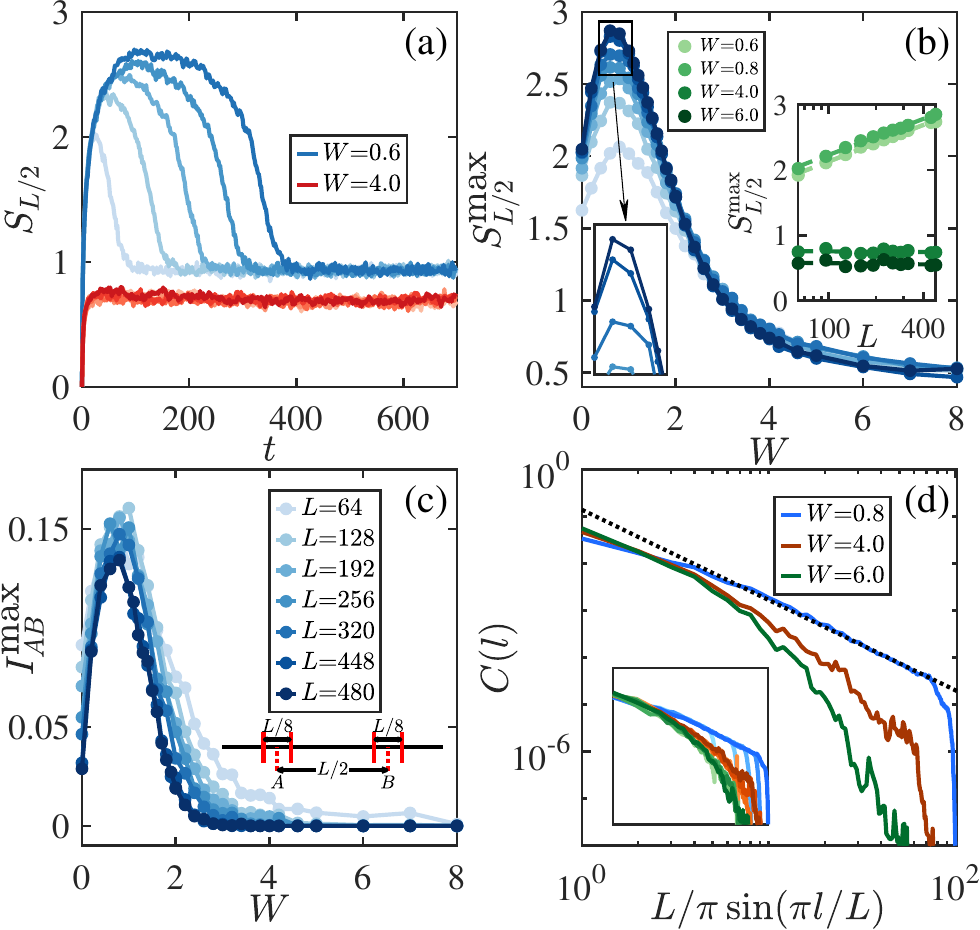}
\caption{(a) The time evolution of the trajectory averaged entanglement entropy $S_{L/2}$ for a
    system under OBCs with $L=64, 128, 192, 256, 320$ (from light to dark colors)
    at $W=0.6$ (blue lines) and $W=4$ (red lines).
    (b) The maximum entanglement entropy $S_{L/2}^{\mathrm{max} }$ and
    (c) the maximum mutual information $I_{AB}^{\mathrm{max}}$ between two subsystems $A$
    and $B$ with respect to $W$ for different system sizes.
    In (b), the inset plots $S^{\mathrm{max}}_{L/2}$ versus $L$ in the linear-log scale at different $W$ with a
    log-law and area-law fit (the dashed lines), and
    in (c), the inset displays the positions and sizes of the two subsystems.
    The mutual information is defined as $I_{AB} = S_A + S_B - S_{A \cup B}$, where $S_A$ and $S_B$ are
    entanglement entropy of the subsystems A and B, respectively, and $S_{A \cup B}$ is the joint
    entanglement entropy.
    In (b) and (c), the maximum value refers to the peak of the corresponding trajectory averaged quantities during time evolution.
    (d) The correlation function $C(l)$ versus $(L/\pi) \sin{(\pi l/L)}$ at $W=0.8, 4, 6$ with $L=320$. The inset displays a data
    collapse for different system sizes $L=64, 128, 192, 256, 320$.}
\label{Fig4}
\end{figure}

\emph{Quasidisorder induced entanglement phase transition}.---
Previous studies have shown that disorder or quasidisorder can drive an entanglement phase transition
in the steady state of the non-Hermitian Hamiltonian dynamics~\cite{Kai2023, ShanZhong2023,LongWen2023}. In our case, we find that the
feedback measurement dynamics under OBCs always leads to a skin state for the steady state even for a large $W$
[see Fig.~\ref{Fig3}(d)], thereby excluding the existence of the phase transition in the steady state.
Furthermore, Fig.~\ref{Fig4}(a) shows that when quasidisorder is sufficiently strong,
$S_{L/2}$ directly approaches a steady value as time evolves, in contrast to the small $W$ case where
the evolution exhibits a dynamical transition, as discussed earlier.
In other words, the dynamical transition with respect to $t/L$ disappears for a large $W$.

%%%

To quantitatively identify the quasidisorder induced disappearance of the dynamical transition,
we calculate the maximum value $S_{L/2}^{\mathrm{max} }$ of the entanglement entropy as time evolves
at each $W$ and plot $S_{L/2}^{\mathrm{max} }$ versus $W$ in Fig.~\ref{Fig4}(b).
The figure shows that $S_{L/2}^{\mathrm{max} }$ first rises to a peak near $W\approx 0.8$ and then collapses to
the same curve, suggesting the entanglement transition from a log-law scaling to
an area-law scaling as $W$ increases [see the inset in Fig.~\ref{Fig4}(b)].
We find that the transition of $S_{L/2}^{\mathrm{max} }$ under OBCs reveals the entanglement
transition in the {\it {steady}} state of the dynamics under {\it PBCs}~\cite{supplement}.
It arises because for weak disorder,
at short times before the
transition to the area-law regime, the dynamics is mainly governed by the system under PBCs.
For strong disorder, since both skin states and localized states respect the
area-law entanglement scaling, the entanglement dynamics under OBCs
shows a similar growth behavior as its dynamics under PBCs.

We further plot the maximum mutual information $I_{AB}^{\mathrm{max}}$ between two subsystems A and B
[see the inset of Fig.~\ref{Fig4}(c)] in the dynamics under OBCs as a function of the quasidisorder strength $W$
in Fig.~\ref{Fig4}(c). Similar to the entanglement entropy, we see that $I_{AB}^{\mathrm{max}}$ first develops a
peak near $W\approx 0.8$ and then declines to zero, providing additional evidence for the quasidisorder
induced transition. We also observe that while $I_{AB}^{\mathrm{max}}$ declines to zero at a smaller $W$
for a larger system, the two lines representing the two largest systems ($L=448,480$) almost overlap, suggesting
a possible smooth transition in the thermodynamic limit.
To gain further insights into this transition, we also calculate the correlation function
$C(l)$ of the evolving state at the time when $S_{L/2}$ reaches its maximum value.
We find that for weak disorder, the correlation function exhibits a power-law decay, whereas for
strong disorder, it shows an exponential decay [see Fig.~\ref{Fig4}(d)].

In summary, we have demonstrated a novel dynamical transition
with respect to a rescaled time in a continuously monitored free fermion
system under OBCs. We show that the
transition arises from the competition between the bulk dynamics and boundary
skin effects.
Our findings are expected to have broader applicability beyond our specific scenario, as other
systems described by the Lindblad master equation can also exhibit the skin effects~\cite{Song2019PRL, Haga2021PRL, Yang2022PRR}.
In addition, we find that while the steady state is still the skin state in the
presence of strong quasidisorder, the maximum entanglement exhibits
a transition from a logarithmic to an area-law scaling as the quasidisorder strength increases.
Since onsite disorder is ubiquitous, we have provided the results for
the case with onsite disorder (rather than quasidisorder) in the Supplemental Material, showing that
the dynamical transition persists.
We note that the dynamical transition also occurs for other initial states~\cite{supplement}.
Our work is also reminiscent of the phenomenon of a metastable state decaying to a steady state in Markovian 
open quantum dynamics~\cite{Juan2016PRL}. However,
the latter does not necessarily correspond to a sharp transition~\cite{supplement}.
Finally, we propose a quantum circuit to observe the novel dynamical transition
in the Supplemental Material~\cite{supplement}, which can be realized with
superconducting qubits~\cite{Wallraff2021RMP}, trapped ions~\cite{Monroe2021RMP} or 
Rydberg atoms~\cite{Saffman2020RMP,Whitlock2021}.

\begin{acknowledgments}
We thank J.-H. Wang, X. Feng, S. Liu, and X. Gao for helpful discussions.
This work is supported by the National Natural Science Foundation of China (Grant No. 11974201),
Tsinghua University Dushi Program
and Innovation Program for Quantum Science and Technology (Grant No. 2021ZD0301604).
We also acknowledge the support by center of high performance computing, Tsinghua University
\end{acknowledgments}

\begin{widetext}
	%%%%%%%%%% Prefix a "S" to all equations, figures, tables and reset the counter %%%%%%%%%%
	\setcounter{equation}{0} \setcounter{figure}{0} \setcounter{table}{0} %
	\renewcommand{\theequation}{S\arabic{equation}} \renewcommand{\thefigure}{S%
		\arabic{figure}}
	%\renewcommand{\bibnumfmt}[1]{[S#1]} \renewcommand{%
	%\citenumfont}[1]{S#1}
	%%%%%%%%%% Prefix a "S" to all equations, figures, tables and reset the counter %%%%%%%%%%
	
    In the Supplemental Material, we will provide details on how to derive and solve the stochastic 
    Schr{\"o}dinger equation in Section S-1, 
    show the probability of finding the evolving state in the ideal many-body skin state for a system
    with quasidisorder in Section S-2,
    discuss the results of the dynamics of the system under PBCs in Section S-3,
    present the results of the dynamics of the system with onsite disorder in Section S-4,
    show that the dynamical transition also occurs for other initial states in Section S-5,	
    compare the dynamical transition with the phenomenon of a metastable state
    decaying to a steady state in Section S-6, and finally propose an experimental scheme
    to observe the dynamical transition in Section S-7.
    
    \section{S-1. Quantum Jump Evolution}
    In the section, we will follow Refs.~\cite{QMCbook,ChenFang2022arXiv,ShuChen2022arXiv} to derive the stochastic Schr{\"o}dinger equation for 
    a continuously monitored system and present how we solve the equation to obtain the trajectories 
    for a comprehensive reading. In addition, we provide details on how to calculate the average velocity 
    per particle for a system under OBCs.
    
    \subsection{A. The stochastic Schr{\"o}dinger equation}
    To derive the stochastic Schr{\"o}dinger equation, we consider a system subject to a unitary evolution 
    of a Hamiltonian $\hat{H}$ and continuous measurements.
    In each infinitesimal time interval $[t, t+dt]$, we first evolve the state of the system $\ket{\psi(t)}$ by 
    the Hamiltonian $\hat{H}$, that is,
    \begin{equation}
    \ket{\psi(t+dt)} = (1 - \ii \hat{H} dt)\ket{\psi(t)}.
    \end{equation}
    Subsequently, we perform a measurement on the state $\ket{\psi(t+dt)}$.
    The measurement is described by the following measurement operators,
    \begin{equation}
        \begin{aligned}
            \hat{M}_m &= \hat{L}_m \sqrt{\gamma d t}\\
            \hat{M}_0 &=\sqrt{ 1 - \gamma \hat{R} dt} = 1 - \frac{\gamma}{2} \hat{R} dt,
        \end{aligned}
    \end{equation}
    where $m=1,2,\dots,M$, $\hat{L}_m$ are quantum jump operators, $\gamma$ is the monitoring rate, and
    \begin{equation}
        \hat{R} = \sum_{m = 1}^{M} \hat{L}_m^\dagger \hat{L}_m.
    \end{equation}
    In the final step, we have neglected the higher-order contributions in $dt$. Clearly, these measurement operators 
    satisfy the completeness equation,
    $\sum_m \hat{M}_m^\dagger \hat{M}_m = 1$.
    
    The post measurement state is governed by a random process. Specifically, 
    through a non-Hermitian evolution, the state will become
    \begin{equation}
        \ket{\psi(t + dt)}
        \rightarrow
        \frac{\hat{M}_0 \ket{\psi (t+dt)}}{\sqrt{\bra{\psi (t+dt)} \hat{M}_0^\dagger \hat{M}_0 \ket{\psi (t+dt)}}}
        =
        \frac{\hat{U}_0 \ket{\psi (t)}}{\sqrt{\bra{\psi (t)} \hat{U}_0^\dagger \hat{U}_0 \ket{\psi (t)}}}
        =
        \left( 1 - \ii \hat{H}_\text{eff} dt + \frac{\gamma}{2} \bra{\psi(t)} \hat{R} \ket{\psi(t)} dt \right) \ket{\psi (t)}
    \end{equation}
    with the probability of
    \begin{equation}
        P(0) = \bra{\psi(t+dt)} \hat{M}_0^\dagger \hat{M}_0 \ket{\psi(t+dt)} = \bra{\psi(t)} \hat{U}_0^\dagger \hat{U}_0 \ket{\psi(t)},
    \end{equation}
    where $\hat{U}_0 = \hat{M}_0 - \ii \hat{H} dt = 1 - \ii \hat{H}_\text{eff} dt$ is the evolution operator
    and $\hat{H}_\text{eff} = \hat{H} - \ii \gamma \hat{R}/2$ is the non-Hermitian effective Hamiltonian.
    Through a jump process, the state can become 
    \begin{equation}
        \ket{\psi(t + dt)}
        \rightarrow
        \frac{\hat{M}_m \ket{\psi (t+dt)}}{\sqrt{\bra{\psi (t+dt)} \hat{M}_m^\dagger \hat{M}_m \ket{\psi (t+dt)}}}
        =
        \frac{\hat{M}_m \ket{\psi (t)}}{\sqrt{\bra{\psi (t)} \hat{M}_m^\dagger \hat{M}_m \ket{\psi (t)}}}
        =
        \frac{\hat{L}_m \ket{\psi (t)}}{\sqrt{\bra{\psi (t)} \hat{L}_m^\dagger \hat{L}_m \ket{\psi (t)}}}
    \end{equation}
    with the probability of
    \begin{equation}
    P(m) = \bra{\psi(t+dt)} \hat{M}_m^\dagger \hat{M}_m \ket{\psi(t+dt)} = \bra{\psi(t)} \hat{M}_m^\dagger \hat{M}_m \ket{\psi(t)}.
    \end{equation}
    
    We now introduce a set of random variables $\{ (dW_1,dW_2,\dots, dW_M): dW_m \in \{ 0,1 \} \text{ with } m=1,2,\dots,M\}$ satisfying 
    $dW_m dW_n = \delta_{mn} dW_m$ (i.e., there is at most one $dW_m = 1$). The probability that $dW_m=1$ is
    $\mathbb{E}(dW_m) = P(m) = \bra{\psi(t)} \hat{L}_m^\dagger \hat{L}_m \ket{\psi(t)} \gamma dt$.
    The random process can be represented by a differential equation,
    \begin{equation}
    \label{next_state}
    \begin{aligned}
    \ket{\psi(t + dt)} &=
    \sum_{m=1}^{M} dW_m
    \frac{\hat{L}_m \ket{\psi (t)}}{\sqrt{\bra{\psi (t)} \hat{L}_m^\dagger \hat{L}_m \ket{\psi (t)}}}
    +
    (1 - \sum_{m=1}^{M} dW_m)
    (1 - \ii \hat{H}_\text{eff} dt + \frac{\gamma}{2} \bra{\psi(t)} \hat{R} \ket{\psi(t)} dt) \ket{\psi (t)}
    \\
    &=
    \left( 1 - \ii \hat{H}_\text{eff} dt + \frac{\gamma}{2} \bra{\psi(t)} \hat{R} \ket{\psi(t)} dt \right) \ket{\psi (t)}
    +
    \sum_{m=1}^{M} dW_m
    \left( \frac{\hat{L}_m}{\sqrt{\bra{\psi (t)} \hat{L}_m^\dagger \hat{L}_m \ket{\psi (t)}}} - 1 \right)
    \ket{\psi (t)},
    \end{aligned}
    \end{equation}
    where we have neglected the higher-order terms in $dt$.
    From Eq.~(\ref{next_state}), we have
    \begin{equation}
    \label{QJ_SSE}
    d\ket{\psi(t)}
    =
    \left( - \ii \hat{H}_\text{eff} + \frac{\gamma}{2} \bra{\psi(t)} \hat{R} \ket{\psi(t)} \right)
    \ket{\psi (t)} dt
    +
    \sum_{m=1}^{M} dW_m
    \left( \frac{\hat{L}_m}{\sqrt{\bra{\psi (t)} \hat{L}_m^\dagger \hat{L}_m \ket{\psi (t)}}} - 1 \right)
    \ket{\psi (t)},
    \end{equation}
    which is the stochastic Schr{\"o}dinger equation in the main text.
    The constant term $\frac{\gamma}{2} \bra{\psi(t)} \hat{R} \ket{\psi(t)}$ can be neglected if we enforce $\langle \psi(t) | \psi(t) \rangle = 1$.

    \subsection{B. Numerical simulation of the stochastic Schr{\"o}dinger equation}
    For an efficient numerical simulation of the stochastic Schr{\"o}dinger equation~[Eq.~(\ref{QJ_SSE})], 
    we use a finite time interval $\Delta t$ and consider the following dynamics.
    Given a state $\ket{\psi(t)}$ at time $t$, we first evolve the state using the non-Hermitian effective 
    Hamiltonian $\hat{H}_\text{eff}$, obtaining
    \begin{equation}
    \label{numerical_evolution}
    \ket{\psi(t+\Delta t)} = \frac{1}{\mathcal{N}} e^{-\ii \hat{H}_\text{eff} \Delta t} \ket{\psi(t)}.
    \end{equation}
    Considering the fact that there can be multiple quantum jumps when the time interval $\Delta t$ is finite, 
    the measurement process is approximated by~\cite{ChenFang2022arXiv,ShuChen2022arXiv}
    \begin{equation}
    \label{numerical_jump}
    \ket{\psi(t+\Delta t)} \rightarrow \frac{1}{\mathcal{N}'} \prod_{m} \left( \Delta W_m \hat{L}_m +1- \Delta W_m \right) \ket{\psi(t+\Delta t)},
    \end{equation}
    with $\Delta W_m \in \{ 0,1 \}$ and $P(\Delta W_m = 1) = \bra{\psi(t)} \hat{L}_m^\dagger \hat{L}_m \ket{\psi(t)} \gamma \Delta t$ (there can be multiple $\Delta W_m=1$).
    Here $\mathcal{N}$ and $\mathcal{N}'$ are normalization constants to ensure that the states are normalized.
    We note that when $\Delta t \rightarrow 0$,
    the probability for two or more quantum jumps, which is $O(\Delta t^2)$, is negligible compared to the probability for one quantum jump $O(\Delta t)$,
    recovering the original quantum jump evolution~[Eq.~(\ref{next_state})].
    
    Since both the non-Hermitian evolution~Eq.~(\ref{numerical_evolution}) and the quantum jumps~Eq.~(\ref{numerical_jump}) preserve 
    the Slater determinant form, we can represent the state at time $t$, $\ket{\psi(t)}$ by an $L \times N$ matrix $U(t)$, 
    with $L$ being the system size and $N$ being the particle number.
    Each column of $U(t)$ corresponds a single-particle state, and the many-body state $\ket{\psi(t)}$ is given by
    \begin{equation}
    \ket{\psi(t)} = \prod_{n=1}^{N} \left( \sum_{i=1}^L  [U(t)]_{in} \hat{c}_i^\dagger \right) \ket{0}.
    \end{equation}
    Suppose that the non-Hermitian effective Hamiltonian is written as
    \begin{equation}
    \hat{H}_\text{eff} = \sum_{i,j=1}^{L} [H_\text{eff}]_{ij} \hat{c}_i^\dagger \hat{c}_j,
    \end{equation}
    then the non-Hermitian evolution~Eq.~(\ref{numerical_evolution}) can be efficiently simulated by
    \begin{equation}
    U(t+\Delta t) = \text{qr}[e^{- \ii H_\text{eff} \Delta t} U(t)],
    \end{equation}
    where $\text{qr}$ stands for the QR decomposition to make sure $\ket{\psi(t + \Delta t)}$ is normalized.
    
    Now, we focus on the simulation of the quantum jump $\ket{\psi} \rightarrow \hat{L}_l \ket{\psi}/\sqrt{ \bra{\psi} \hat{L}_l^\dagger  \hat{L}_l  \ket{\psi}}$.
    The jump operator is given by
    \begin{equation}
    \hat{L}_l = e^{\ii \pi \hat{n}_{l+1} } \hat{d}_l^\dagger \hat{d}_l,
    \end{equation}
    with $\hat{d}_l^\dagger = \sum_{i=1}^L [d_l]_i  \hat{c}_i^\dagger$ and $d_l$ being a column vector specified by $[d_l]_i = \frac{1}{\sqrt{2}}(\delta_{i,l} - \ii \delta_{i,l+1})$.
    Let
    \begin{equation}
    \hat{\gamma}_n^\dagger = \sum_{i=1}^L U_{in} \hat{c}_i^\dagger,
    \end{equation}
    we have
    \begin{equation}
    \ket{\psi} = \hat{\gamma}_1^\dagger \hat{\gamma}_2^\dagger \dots \hat{\gamma}_N^\dagger \ket{0},
    \end{equation}
    and
    \begin{equation}
    \hat{d}_l^\dagger \hat{d}_l \ket{\psi} =
    \langle d_l | \gamma_1 \rangle \hat{d}_l^\dagger \hat{\gamma}_2^\dagger \dots \hat{\gamma}_N^\dagger \ket{0}
    +
    \langle d_l | \gamma_2 \rangle \hat{\gamma}_1^\dagger \hat{d}_l^\dagger \hat{\gamma}_3^\dagger \dots \hat{\gamma}_N^\dagger \ket{0}
    + \dots +
    \langle d_l | \gamma_N \rangle \hat{\gamma}_1^\dagger \hat{\gamma}_2^\dagger \dots \hat{\gamma}_{N-1}^\dagger \hat{d}_l^\dagger \ket{0},
    \end{equation}
    with $\ket{d_l} = \hat{d}_l^\dagger \ket{0}$ and $\ket{\gamma_n} = \hat{\gamma}_n^\dagger \ket{0}$.
    Without loss of generality, we assume $\langle d_l | \gamma_1 \rangle \neq 0$.
    To simplify this equation, we can perform the following transformation,
    \begin{equation}
    \hat{\gamma}_n^\dagger \rightarrow \hat{\gamma}_n'^\dagger = \hat{\gamma}_n^\dagger +
     \left(\delta_{n1} - \frac{\langle d_l | \gamma_n \rangle}{\langle d_l | \gamma_1 \rangle} \right) \hat{\gamma}_1^\dagger.
    \end{equation}
    Equivalently, we have $U \rightarrow U'$ with
    \begin{equation}
    U_n \rightarrow U_n' = U_n + \left(\delta_{n1} - \frac{\langle d_l | \gamma_n \rangle}{\langle d_l | \gamma_1 \rangle}  \right) U_1,
    \end{equation}
    where $U_n$ ($U_n'$) is the $n$-th column of $U$ ($U'$).
    We note that such transformation does not change the many-body state $ \ket{\psi}$ given that $(\hat{\gamma}_1^\dagger)^2 = 0$.
    After the transformation, we have $\langle d_l | \gamma_n' \rangle = 0$ for $n \neq 1$, so
    \begin{equation}
    \hat{d}_l^\dagger \hat{d}_l \ket{\psi} =
    \langle d_l | \gamma_1' \rangle \hat{d}_l^\dagger \hat{\gamma}_2'^\dagger \dots \hat{\gamma}_N'^\dagger \ket{0}
    =
    \langle d_l | \gamma_1' \rangle
    \prod_{n=1}^{N} \left( \sum_{i=1}^L  \tilde{U}_{in} \hat{c}_i^\dagger \right) \ket{0}
    \end{equation}
    with
    \begin{equation}
    \tilde{U} =
    \begin{pmatrix}
    d_l & U_2' & U_3' & ... & U_n'
    \end{pmatrix}.
    \end{equation}
    We finally arrive at
    \begin{equation}
    \frac{\hat{L}_l \ket{\psi}}{\sqrt{ \bra{\psi} \hat{L}_l^\dagger  \hat{L}_l  \ket{\psi}}}
    =
    \prod_{n=1}^{N} \left( \sum_{i=1}^L  \tilde{U}'_{in} \hat{c}_i^\dagger \right) \ket{0}
    \end{equation}
    with $\tilde{U}' = \text{qr}(e^{\ii \pi M_{l+1}} \tilde{U})$ and $[M_{l+1}]_{ij} = \delta_{i,l+1} \delta_{j,l+1}$.
    
    \subsection{C. The average velocity $v(t)$ per particle under OBCs}\label{Supp:velocity}
    In this subsection, we will show how to calculate the average velocity $v(t)$ per particle for a system
    under OBCs. The velocity under OBCs is defined as
    \begin{equation}\label{Eq: def}
        v(t)=\frac{d}{dt}[ \langle\hat{x}\rangle ]=\frac{2}{L}\sum_l l\frac{d}{dt} [\langle\hat{n}_l\rangle ].
    \end{equation}
    Here, $[\langle\hat{n}_l\rangle ]=\textrm{Tr}(\rho_t \hat{n}_l)$ ($[\dots]$ denotes the trajectory average) 
    where $\rho_t$ is the density matrix
    that evolves based on the master equation
    \begin{equation} \label{Eq:LMaster}
        \frac{d\rho_t}{dt}=-i\hat{H}_{\textrm{eff}}\rho_t+i \rho_t \hat{H}_{\textrm{eff} }^\dagger+
        \gamma \sum_m \hat{L}_m \rho_t \hat{L}_m^\dagger.
    \end{equation}
    Thus, we obtain
    \begin{equation}\label{Eq: Master}
        \frac{d}{dt}[\langle\hat{n}_l\rangle ]=-i [\langle \hat{n}_l\hat{H}_{\text{eff}}\rangle ]+
        i [\langle \hat{H}_{\text{eff}}^\dagger\hat{n}_l\rangle ]+
        \gamma \sum_m [ \langle \hat{L}_m^\dagger\hat{n}_l\hat{L}_m\rangle ].
    \end{equation}
    Since each trajectory has the Slater determinant form, we can apply the Wick's theorem, e.g.,
    $\langle\hat{c}_i^\dagger \hat{c}_j^\dagger \hat{c}_m \hat{c}_n\rangle=\langle\hat{c}_i^\dagger\hat{c}_n\rangle \langle\hat{c}_j^\dagger \hat{c}_m\rangle -\langle\hat{c}_i^\dagger\hat{c}_m\rangle \langle\hat{c}_j^\dagger \hat{c}_n\rangle$ and write $v(t)$ in terms of two-point correlation functions,
    \begin{equation}
        v(t)=\frac{2}{L}\sum_{l=1}^L \left \{\ii l\left [\langle \hat{H}_{\textrm{eff}}^\dagger\hat{n}_l\rangle-\langle \hat{n}_l\hat{H}_{\textrm{eff}}\rangle \right ]+l\sum_m\left [ \frac{1}{2}(\delta_{lm}+\delta_{l,m+1})\gamma\langle\hat{d}^\dagger_m \hat{d}_m\rangle-\langle\hat{n}_l\rangle \ii\langle\hat{H}^\dagger_{\textrm{eff}}-\hat{H}_{\textrm{eff}} \rangle
        -\gamma \langle\hat{d}^\dagger_m\hat{c}_l \rangle\langle\hat{c}_l^\dagger \hat{d}_m \rangle \right]\right \},
    \end{equation}
    where the term $\langle\hat{H}_{\textrm{eff}}^\dagger\hat{n}_l\rangle$ can be calculated by 
    \begin{equation}
    \sum_{i,j}\left [H_{\textrm{eff}}^\dagger \right]_{ij}\langle \hat{c}^\dagger_i\hat{c}_j\hat{n}_l\rangle
    =\sum_{i,j}\left [H_{\textrm{eff}}^\dagger \right]_{ij}\left (\langle \hat{c}^\dagger_i\hat{c}_j\rangle\langle\hat{n}_l\rangle-\langle\hat{c}^\dagger_i\hat{c}_l \rangle\langle\hat{c}^\dagger_l\hat{c}_j \rangle +\delta_{jl}\langle\hat{c}^\dagger_i\hat{c}_l \rangle\right ).
    \end{equation}

    \section{S-2. Dynamical transition with the quasiperiodic potential}\label{Supp:dpt}
    In the main text, we have demonstrated the dynamical transition in the presence of quasidisorder by showing the change in
    the entanglement scaling as time evolves. In this subsection, we 
    present the probability of the evolving state $|\psi_t\rangle$ being in the ideal many-body skin state, that is, 
    $f_{\textrm{skin}}=[|\langle\psi_{\text{skin}}|\psi_t\rangle|^2]$,
    for different system sizes when the quasidisorder strength $W=1$ in Fig.~\ref{figs1}.  
    The figure illustrates that the probability suddenly rises across the transition point, similar to the case without disorder.
    \begin{figure}
      \includegraphics[width=0.3\linewidth]{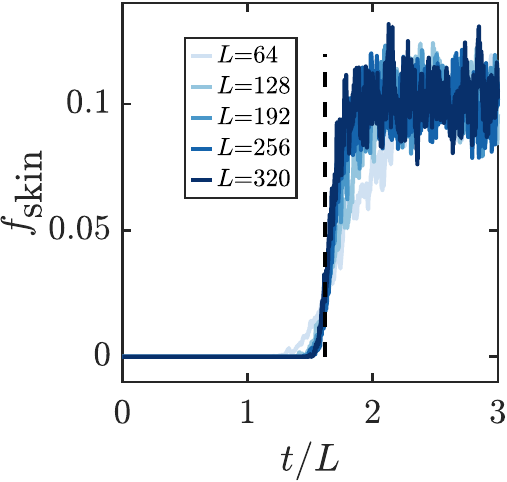}
      \caption{The time evolution of the probability of finding the evolving state $|\psi_t\rangle$ 
            in the many-body skin state $|\psi_{\text{skin} }\rangle$ averaged over trajectories in the presence of
            quasidisorder with $W=1$ ($f_{\textrm{skin}}=[|\langle \psi_{\textrm{skin}}| \psi_t \rangle|^2] $). 
            The black dashed line highlights the dynamical transition point in Fig. 3(b) in the main text.}
      \label{figs1}
    \end{figure}%%%
    
    \section{S-3. The dynamics of the system under PBCs}\label{Supp: EEPBC}
    In the main text, we have shown the existence of the dynamical transition for a system under OBCs 
    that arises from the competition between the bulk dynamics and boundary skin effects. We have also shown 
    that under OBCs, the entanglement entropy first increases and then declines to a steady value as time evolves 
    for weak quasidisorder. We argue that the maximum entanglement entropy with respect to the 
    quasidisoder strength $W$ reveals the entanglement transition of the steady state of the dynamics
    under PBCs. In this section, we will provide the results of the dynamics under PBCs in Fig.~\ref{figs2} to 
    support the argument. 
    Figure~\ref{figs2}(a) illustrates that as time progresses, the bipartite entanglement entropy rises rapidly, 
    reaching a steady value, in stark contrast to the OBC case [see Fig. 4(a) in the main text]. 
    The dynamical transition thus does not happen in the PBC case, which is consistent to the 
    fact that the dynamical transition appears due to the competition between the bulk dynamics and 
    boundary skin effects. 
    
    However, for the steady state, the numerical results suggest a smooth transition of
    the entanglement entropy of the {\it steady} state from a log-law to an area-law scaling with respect to $W$ as 
    shown in Fig.~\ref{figs2}(b) (see the inset for the change in the scaling), similar to the OBC case. In addition, the correlation function $C(l)$
    exhibits a power-law decay at weak quasidisorder and develops into the exponential decay 
    at strong quasidisorder, which agrees well with the OBC case. These results strongly suggest 
    that the transition of the maximum entanglement entropy present in the OBC case arises from the initial bulk dynamics.  
    \begin{figure}
      \includegraphics[width=0.6\linewidth]{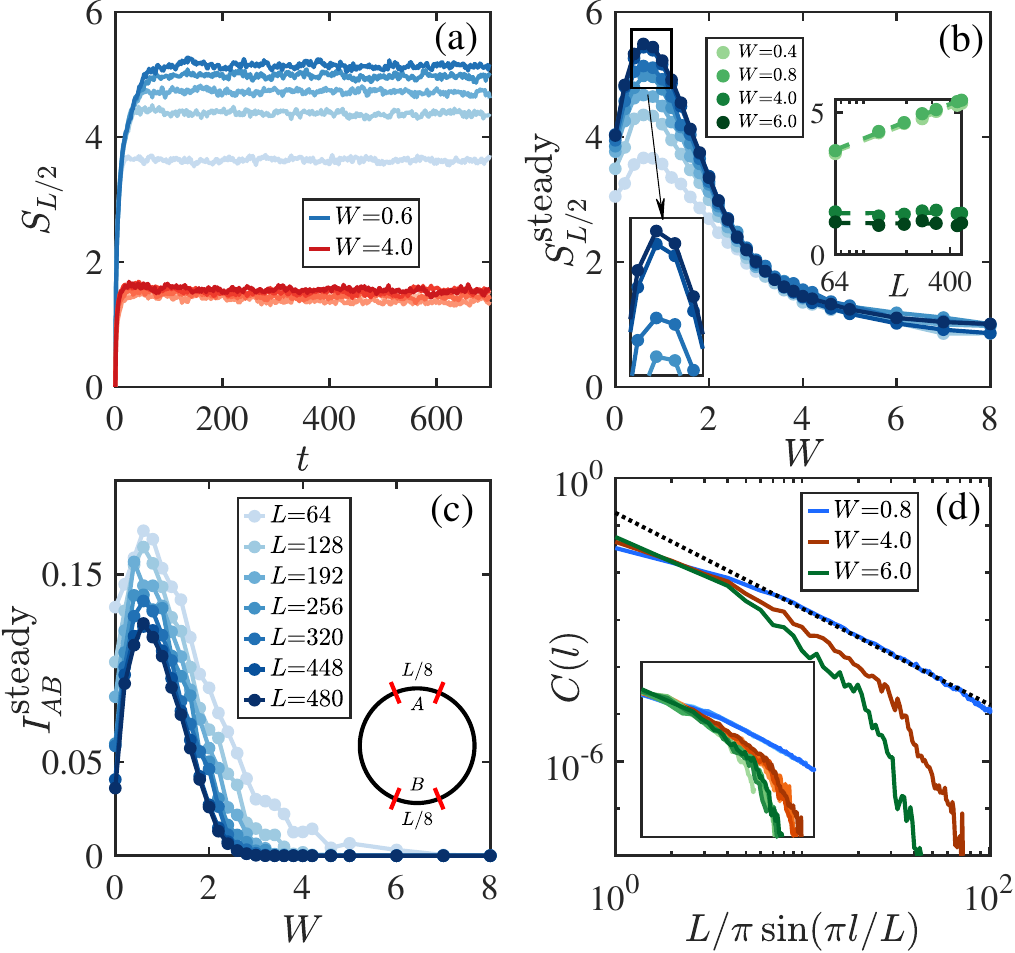}
      \caption{The dynamics of the system under PBCs. 
          (a) The time evolution of the trajectory averaged bipartite 
          entanglement entropy $S_{L/2}$ for system sizes $L=64, 128, 192, 256, 320$ (light to dark colors) 
          at $W=0.6$ (blue lines) and $W=4$ (red lines). 
          (b) The steady-state value of $S_{L/2}$ with respect to the quasidisorder strength $W$ for different system sizes
          $L=64, 128, 192, 256, 320, 448, 480$ (from light to dark colors).
          Inset: the linear-log plot of the steady-state value of $S_{L/2}$ versus $L$ at different $W$.
          (c) The trajectory averaged mutual information between two subsystems $A$ and $B$ (see the inset)
          for the steady state with respect to $W$.
          (d) The correlation function $C(l)$ of the steady state with respect to $(L/\pi)\sin{(\pi l/L)}$ for different $W$, 
          with $L=320$. The inset shows the data collapse of $C(l)$ for $L=64, 128, 192, 256, 320$.}
      \label{figs2}
    \end{figure}%%%
    
    \section{S-4. The dynamical transition of a system with onsite disorder} \label{Supp:Anderson}
    In the main text, we have studied the dynamics of a system with quasidisorder. Since onsite disorder is ubiquitous, 
    in this section, we will provide the results for a system with onsite disorder. To study the effects of 
    onsite disorder, we introduce the term 
    $\sum_l m_l \hat{c}_l^\dagger \hat{c}_l$ in the Hamiltonian (1) in the main text,
    where $m_i$ represents the onsite disorder that is uniformly sampled in $[-W/2,W/2]$ ($W$ now denotes
    the disorder strength rather than the quasidisorder one). In our numerical calculations, we average our
    results over 224-1000 sample configurations for one trajectory. 
    In Fig.~\ref{figs3}, we map out the phase diagram
    of the bipartite entanglement entropy with respect to $t/L$ and $W$, showing the existence of a dynamical 
    transition, similar to the case with quasidisorder.
    To explicitly illustrate the transition from the log-law growth to an area-law behavior, we plot the time 
    evolution of $S_{L/2}$ in Fig.~\ref{figs3}(b). It exhibits an increase-decrease-steady pattern, which closely 
    resembles the quasidisorder case. Additionally, we present the scaling of $S_{L/2}$ before the transition 
    point in Fig.~\ref{figs3}(c) to demonstrate that $S_{L/2}$ follows the log-law growth in that regime.  
    \begin{figure}
      \includegraphics[width=0.9\linewidth]{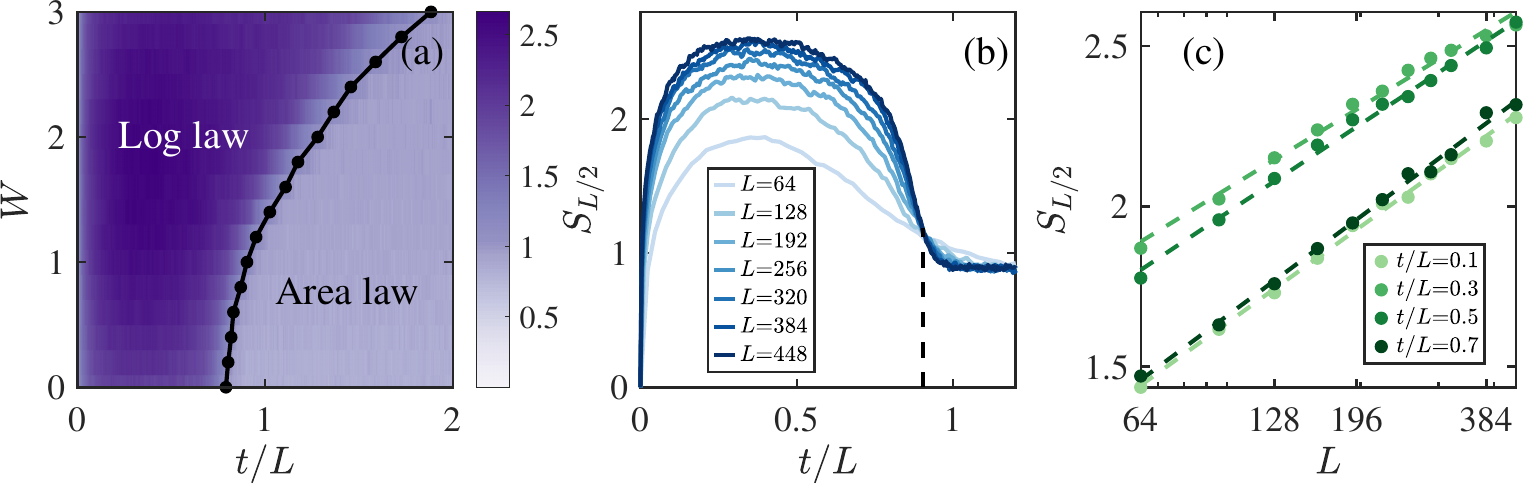}
      \caption{The dynamics of the system with onsite disorder under OBCs.
          (a) The phase diagram of the trajectory averaged bipartite entanglement entropy 
          with respect to $W$ and $t/L$ for the system with $L=320$. 
          The black dotted line represents the dynamical transition line, separating the log-law regime and 
          the area-law regime. 
          (b) The trajectory averaged bipartite entanglement entropy $S_{L/2}$ as a function of $t/L$ for different 
          system sizes at $W=1$. The black dashed line marks out the dynamical transition point. 
          (c) The scaling of $S_{L/2}$ with respect to $L$ in a linear-log scale with $L$ up to 448. }
      \label{figs3}
    \end{figure}%%%
    
    Furthermore, similar to the quasidisorder case, our numerical results suggest the existence of a transition for 
    the maximum entanglement entropy $S_{L/2}^\textrm{max}$ from the log-law regime to the area-law regime 
    with respect to the disorder strength
    in the disordered system as shown in Fig.~\ref{figs4}.
    
    \begin{figure} 
      \includegraphics[width=0.6\linewidth]{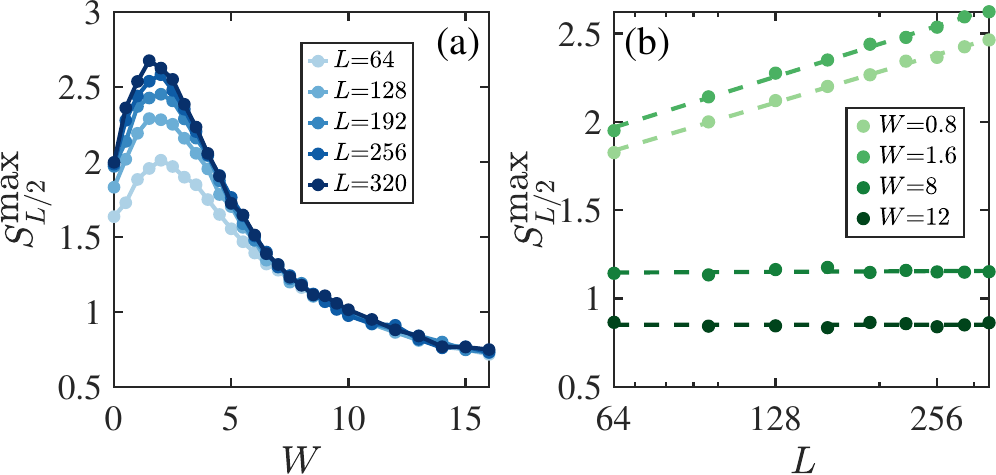}
      \caption{(a) The maximum entanglement entropy with respect to the disorder strength $W$
          for different system sizes under OBCs. 
          (b) The scaling of $S_{L/2}^\textrm{max}$ with $L$ for several disorder strengths.}
      \label{figs4}
    \end{figure}%%%
    
    \section{S-5. Dynamical transition for other initial states}
    In the main text, we have shown that the dynamical transition can happen when the initial state is prepared as
    the Ne\' {e}l state. In this section, we will choose the ground state of the fermion chain in Eq. (1) at half filling 
    and randomly half-filled Fock states as initial states for time evolution to demonstrate that the dynamical transition
    also occurs for these initial states.
    
    In the first two rows of Fig.~\ref{Fig-initial}, we plot the results obtained by choosing the ground state of the fermion
    chain as the initial state. We see that for both $W=0$ (without quasidisorder) and $W=1$ (with quasidisorder),
    the dynamical transition appears with the transition point close to that for the case with the Ne\' {e}l state
    taken as an initial state. In the third row, we display the results obtained by choosing a randomly half-filled Fock state,
    wherein $L/2$ sites are randomly occupied, as the initial state. The results are averaged over 
    many different random initial configurations. The figure illustrates the existence of the dynamical transition
    with the transition point close to that for the case with the Ne\' {e}l state as the initial state. These results indicate
    that the dynamical transition is not specific to the Ne\' {e}l state as the initial state. 
    
    \begin{figure} 
        \includegraphics[width=1\linewidth]{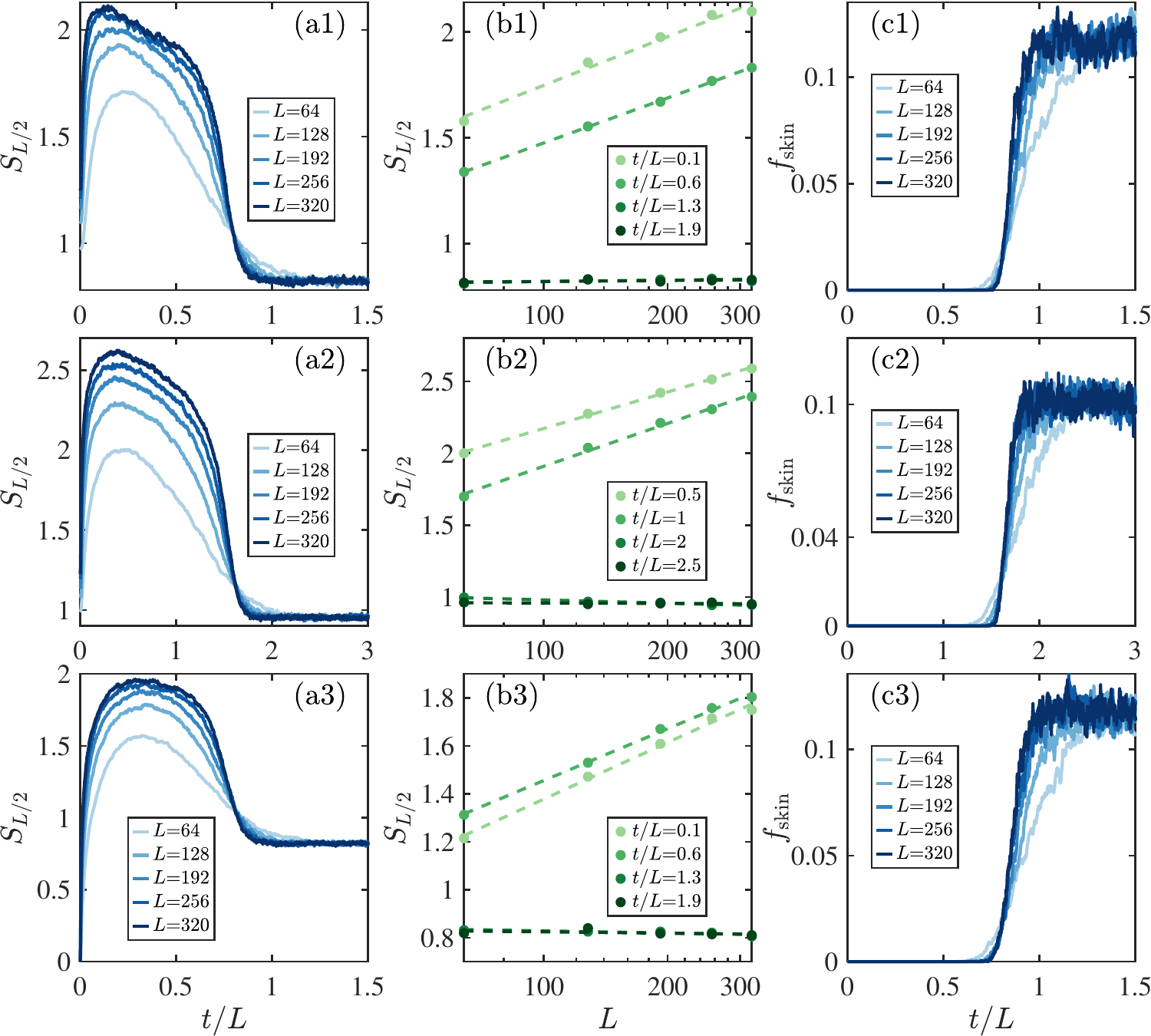}
        \caption{(a1)-(a3) The time evolution of the trajectory averaged bipartite entanglement entropy $S_{L/2}$.
            (b1)-(b3) The linear-log plot of $S_{L/2}$ with respect to the system size $L$, illustrating the log-law
            scaling before the transition and the area-law scaling after the transition.
            (c1)-(c3) The time evolution of the probability $f_{\textrm{skin}}$. 
            In the first and second rows, we choose the ground state of the fermion chain in Eq. (1) at half filling 
            as the initial state for the evolution. 
            In the third row, we choose a random Fock state at half filling as an initial state. The obtained results
            are averaged over 2000 random Fock states as an initial state.    
            In the first and third rows, $W=0$ and in the second one, $W=1$.
            }
        \label{Fig-initial}
    \end{figure}%%%
    
    \section{S-6. Comparison with the traditional metastable state decaying to a steady state} \label{Supp:Metastability}
    In this section, we will explore a possible explanation for the dynamical transition by analyzing the spectrum of a
    Liouvillian superoperator~\cite{Juan2016PRL} and clarify that the phenomenon of a metastable 
    state decaying to a steady state does not necessarily correspond to a transition. 
    
    \begin{figure} 
        \includegraphics[width=1\linewidth]{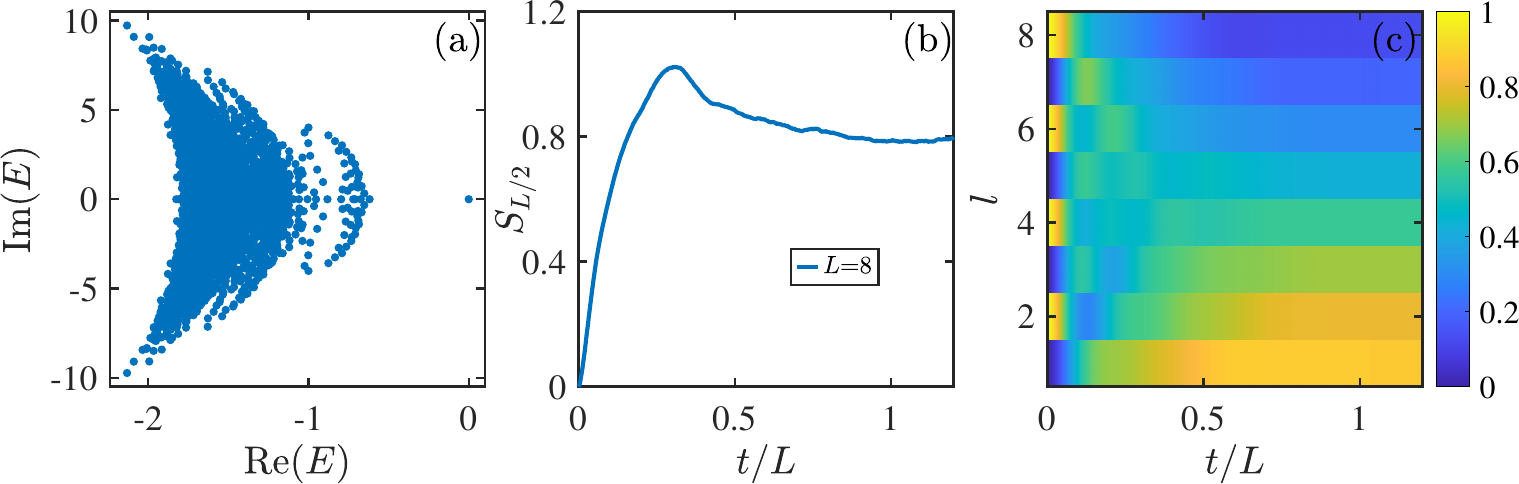}
        \caption{(a) The spectrum of the Liouvillian superoperator $\mathcal{L}$ 
            for the master equation in Eq.~(\ref{Eq:LMaster}) (here we consider OBCs). 
            (b) The time evolution of the trajectory averaged bipartite entanglement entropy $S_{L/2}$ and 
            (c) the trajectory averaged density distribution.
            Here $L=8$.}
        \label{figs5}
    \end{figure}%%%
    
    We first follow Ref.~\cite{Juan2016PRL} to review the fact that the metastability can arise from a large separation 
    between some low-lying eigenvalues and the rest of the spectrum. Specifically, consider the Lindbladian master 
    equation,
    \begin{equation}
        \frac{d\rho_t}{dt}=\mathcal{L}\rho_t,
    \end{equation}
    where $\rho_t$ is the evolving density matrix, and $\mathcal{L}$ is the generator of the dynamics 
    (also called Liouvillian superoperator or Lindbladian). One can easily obtain
    \begin{equation}
        \rho_t=\rho_{\text{SS}}+\sum_{j=1}^{M}c_j e^{\lambda_j t} \rho_j^R,
    \end{equation}
    where $\rho_{\text{SS}}$ is the steady state, $\rho_j^R$ is a right eigenstate of $\mathcal{L}$ 
    corresponding to an eigenvalue $\lambda_j$, that is, $\mathcal{L}\rho_j^R=\lambda_j \rho_j^R$, $c_j$ 
    denotes the weight of the initial state in the right eigenstate $\rho_j^R$, and $M$ denotes the square of 
    the dimension of the Hilbert space. For convenience, we sort the eigenvalues so that $|\text{Re}(\lambda_{n+1})|\ge |\text{Re}(\lambda_{n})|$. 
    
    If there exists a large separation between some low-lying eigenvalues and the rest of the spectrum, i.e., there exists 
    an $m \ge 2$ such that $|\text{Re}(\lambda_{m+1})| \gg |\text{Re}(\lambda_{m})|$,
    then a metastable state occurs in the range $\tau^{\prime\prime} \ll t \ll \tau^\prime$ with $\tau^{\prime\prime} \sim 1/|\text{Re}(\lambda_{m+1})|$
    and $\tau^\prime \sim 1/|\text{Re}(\lambda_{m})|$. This happens because in the range the states with $n>m$ almost decay 
    to zero so that the dynamics is approximately dictated by
    \begin{equation}
        \rho_t=\rho_{\text{SS}}+\sum_{j=1}^{m}c_j e^{\lambda_j t} \rho_j^R.
    \end{equation}
    
    To examine whether the metastable state occurs due to the existence of the energy separation in our case, 
    we plot the spectrum of $\mathcal{L}$ for the master equation in Eq.~(\ref{Eq:LMaster}) (here we consider OBCs)
    in Fig.~\ref{figs5}(a). 
    Note that since $\mathcal{L}$ does not correpond to the single-particle case, we can only diagonalize it for small system sizes 
    (here we take $L=8$). We see that except a large spectral gap, there does not appear a clear separation in the rest part of the spectrum.
    Interestingly, we can still observe an initial increase followed by a decline in the entanglement entropy [Fig.~\ref{figs5}(b)], 
    as well as a particle transport to the lower region in the density distribution [Fig.~\ref{figs5}(c)].
    The results may suggest the presence of a metastable state within a small region in the system, indicating that the metastable state 
    may not be a consequence of a significant gap between low-lying eigenvalues and the remainder of the spectrum. 
    However, this is not conclusive as 
    both the system size and the metastable region are small. We cannot preclude the possibility that a larger system
    (e.g. $L=448$) may display a substantial separation in the spectrum.   
    In fact, in our case, we can explain the existence of the metastable state as arising from the dynamics under PBCs.
    This method also allows for an accurate prediction of the transition point without relying on spectrum calculations 
    that are limited to very small systems (see the discussion in the main text). 
    
    We would also like to clarify that according to Ref.~\cite{Juan2016PRL}, a metastable state can occur in a very small system, 
    implying that the phenomenon of a metastable state decaying to a steady state does not necessarily correspond to a phase transition. 
    For example, the authors in the paper take a three-level system and a two-qubit system to demonstrate the existence of a metastable 
    state due to the presence of a large separation between the low-lying eigenvalues and the rest of the spectrum. While one can 
    observe the decay of the metastable state to a steady one as time evolves, this decay is not a phase transition given that the 
    system sizes are too small.
    
    In our case, we consider the dynamics for different system sizes and our finite-size analysis clearly suggests an existence 
    of a ``phase" transition of the bipartite entanglement entropy and the probability $f_{\text{skin}}$ with 
    respect to $t/L$ in the thermodynamic limit. The subtle point is that in the thermodynamic limit that $L\rightarrow \infty$,
    it takes an infinite amount of time to observe the occurrence of the transition.  
    
    \section{S-7. Experimental scheme} \label{Supp:Scheme}
    In this section, we will propose a quantum circuit (inspired by Ref.~\cite{ChenFang2022arXiv}) to 
    experimentally observe the novel dynamical transition. 
    
    \begin{figure} 
        \includegraphics[width=1\linewidth]{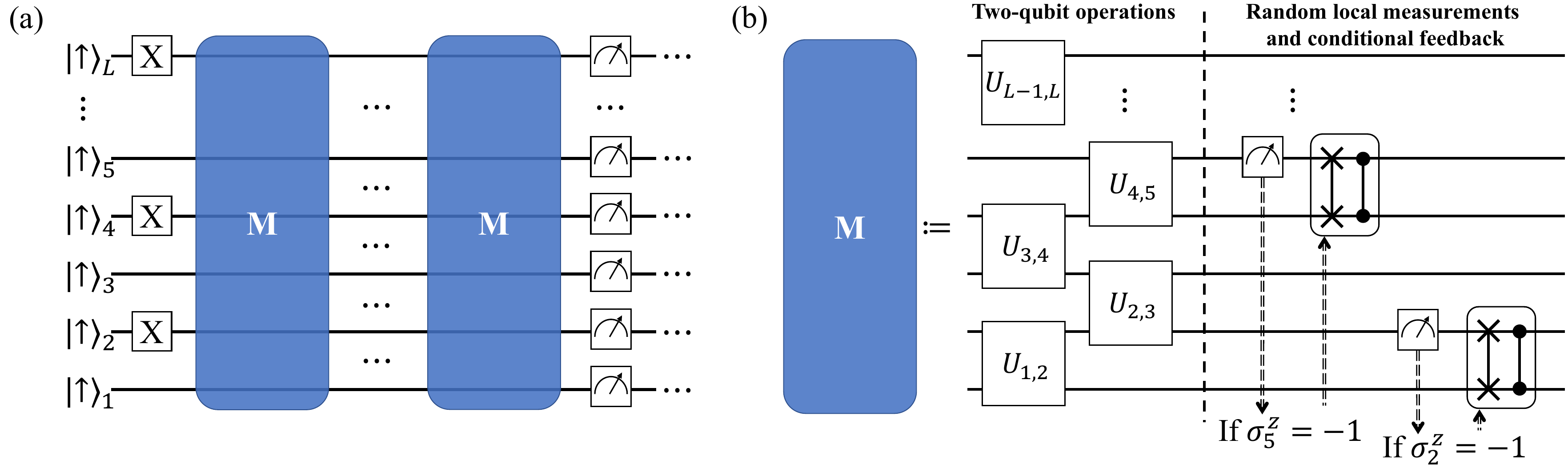}
        \caption{Quantum circuit with $L$ qubits implementing the dynamical transition. 
            (a) The circuit requires the consecutive operation of $l$ modules (labeled by $M$) before
            performing the local measurements at time $t=N_t\delta t$. The module is illustrated in (b), 
            and the detailed explanation can be found in the text. Note that the qubits are labeled from $1$ to $L$ 
            in ascending order from the bottom to the top,  aligning with the site labels in Fig. 1(b) in the main text.}
        \label{figs6}
    \end{figure}%%%
    
    The circuit is shown in Fig.~\ref{figs6}.
    We first initialize the qubits to the state $|\uparrow \uparrow \dots \uparrow\rangle$.
    Next, we apply local $\sigma^x$ gates to the even sites to get the Ne{\'e}l state 
    $\left|\uparrow \downarrow \uparrow \downarrow \dots \uparrow \downarrow \right\rangle$ 
    corresponding to the initial state $|\psi_0\rangle =\prod_{l=1}^{L/2}\hat{c}_{2l}^\dagger |0\rangle$ in the fermion model. 
    Here we use the convention that $|0\rangle \coloneq \left| \uparrow \right\rangle$ and $|1\rangle \coloneq \left| \downarrow \right\rangle$.
    To measure an observable, such as the spin distribution, at time $t=N_t\delta t$ ($N_t$ is a positive integer),
    a consecutive operation of $l$ modules (labeled by $M$) are performed [see Fig.~\ref{figs6}(a)].
    
    Each module contains two parts as shown in Fig.~\ref{figs6}(b).
    In the first part, we apply two-qubit unitary operations in a bricklayer pattern. Each operation realizes 
    the following unitary evolution between two neighboring qubits
    \begin{equation}
        U_{l,l+1}=e^{-i \delta t (\sigma_l^x \sigma_{l+1}^x+\sigma_l^y \sigma_{l+1}^y) },
    \end{equation}
    where $\sigma_l^x$ and $\sigma_l^y$ represent the Pauli matrices acting on the $l$th qubit, 
    and $\delta t$ takes a small real value representing the time step. If we transform the spin model to a 
    fermion model by the Jordan-Wigner transformation, then we will find that these unitary gates realize the 
    evolution of a free fermion model with nearest-neighbor hopping.
    
    In the second part, we perform local $\sigma_l^z$ measurements on each qubit with a probability of $p$. If the measurement 
    outcome on the $l$th qubit is $-1$ in the basis of $\{ \left|\uparrow\right\rangle, \left|\downarrow\right\rangle \}$ (or equivalently $1$ in the 
    basis of $\{ \ket{0}, \ket{1} \}$), then we immediately apply a SWAP gate (indicated by the connected diagonal crosses) 
    followed by a CZ gate (indicated by the connected solid circles) between the $l$th and $(l-1)$th qubits. 
    Such operations realize the feedback function. In the language of the fermion model, the operations transport an 
    electron to the $(l-1)$th site once find its existence at the $l$th site.
    
    \begin{figure} 
        \includegraphics[width=1\linewidth]{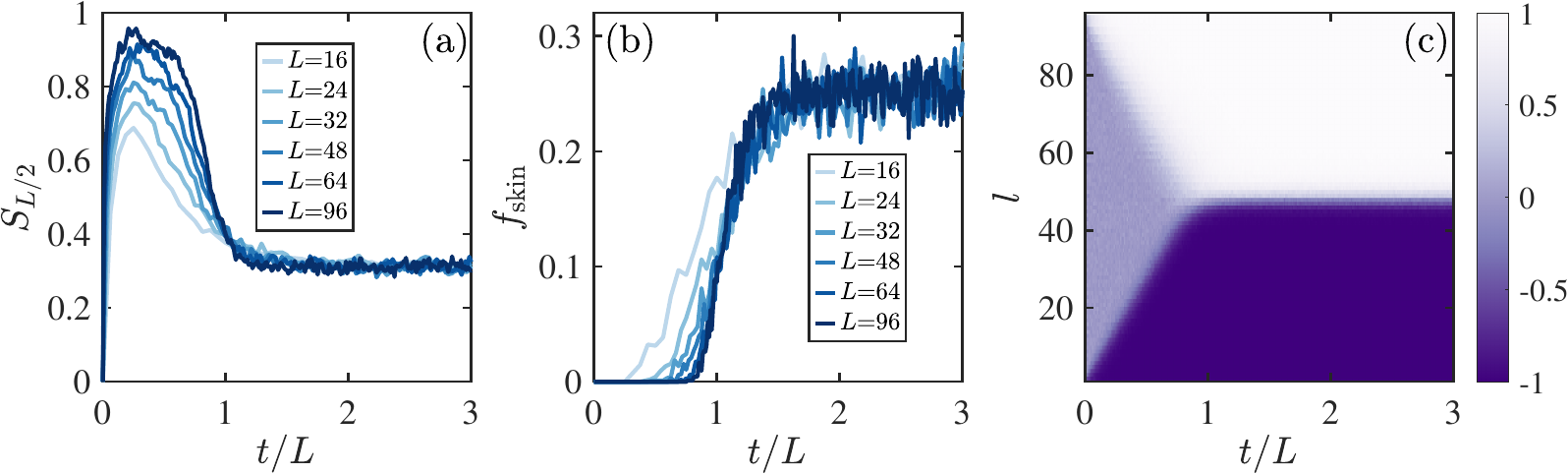}
        \caption{The results for the quantum circuit shown in Fig.~\ref{figs6}. 
            (a) The trajectory averaged bipartite entanglement entropy $S_{L/2}$ 
            and (b) the single-shot measurement averaged probability $f_{\text{skin}}$ of the state 
            $\left|\psi_t\right\rangle$ being in the skin state 
            $\left|\psi_{\text{skin}}\right\rangle=\left|\downarrow \downarrow \dots\downarrow \uparrow\uparrow\dots\uparrow \right\rangle$ 
            with respect to $t/L$ for various system sizes $L$. 
            (c) The time evolution of the single-shot measurement averaged spin distribution for a system of size $L=96$. 
            Here we take $\delta t=0.5$, $p=0.7$ and $N=900$.}
        \label{figs7}
    \end{figure}%%%
    
    \begin{figure} 
        \includegraphics[width=1\linewidth]{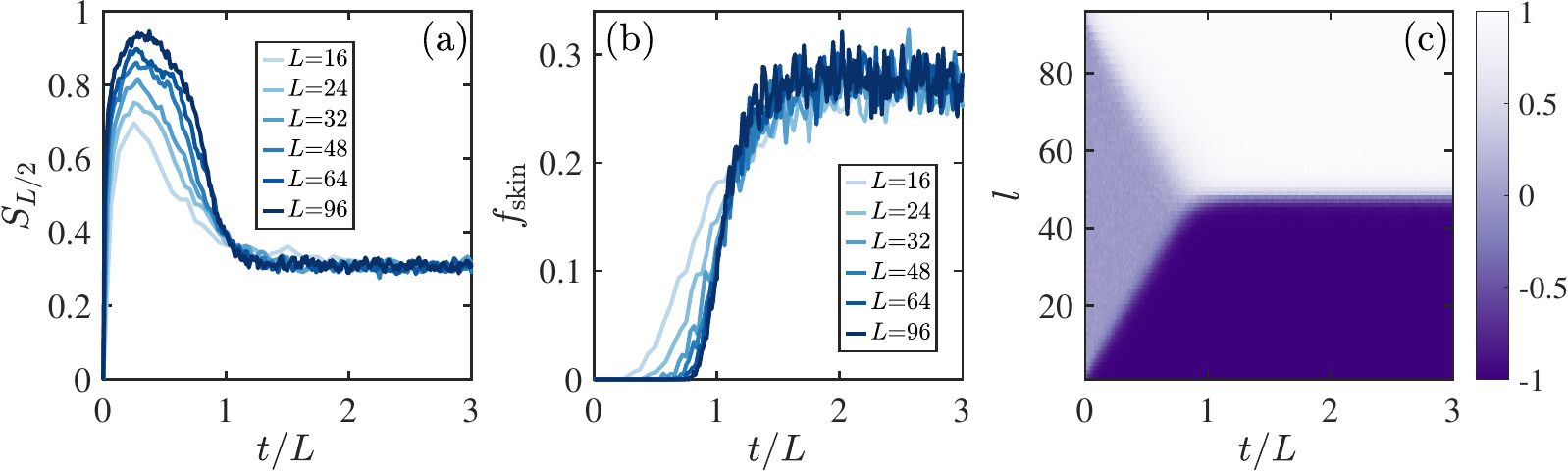}
        \caption{Similar to Fig.~\ref{figs7}, but with quasidisorder at $W=0.25$.}
        \label{figs8}
    \end{figure}%%%
    
    To verify that our circuit works, 
    we compute the entanglement entropy $S_{L/2}$ of the half chain, the overlap $f_{\mathrm{skin}}$ 
    with the skin state $|\psi_{\mathrm{skin} }\rangle=\left|\downarrow \downarrow \dots \downarrow \uparrow \dots \uparrow \uparrow\right \rangle$
    and the spin distribution (corresponding to the density distribution in the fermion model) for the circuit
    at time $t=N_t \delta t$. Before showing the results, we would like to clarify the following points. 
    While in a real experiment, it is very challenging to measure the trajectory averaged entanglement entropy, 
    one can instead probe the spin distribution
    and the overlap $f_{\mathrm{skin}}$ to observe the 
    dynamical transition based on local measurements shown in Fig.~\ref{figs6}(a).
    To measure these quantities, it is not necessary to perform postselection in order to ensure that 
    the state at time $t$ remains fixed for the purpose of measuring their expectation values. 
    In fact, we only need to run the circuit $N$ different times, each for a duration of time $t$, 
    and each time a single-shot local measurement $\{\sigma_l^z\}$ at time $t$ is sufficient. 
    For example, at the $j$th run, if the single-shot local measurement  
    yields a result represented by an array of $-1$ or $1$, that is, 
    $\{\nu_l^j : \nu_l^j\in\{-1,1\},l=1,\dots ,L \}$, then 
    the average spin distribution is given by
    \begin{equation}
        n_l=\frac{\sum_{j=1}^N\nu_l^j}{N}.
    \end{equation}
    Meanwhile, the $f_{\mathrm{skin}}$ is determined by
    \begin{equation}
        f_{\mathrm{skin}}=\frac{\sum_{j=1}^N p_{\mathrm{skin}}^j}{N},
    \end{equation}
    where $p_{\mathrm{skin}}^j \in\{0,1\}$ characterizes whether the measurement result $\{\nu_l^j \}$ 
    is $\{ -1,-1, \dots,-1,1,1, \dots, 1 \}$ 
    corresponding to an ideal skin state. 
    
    Figure~\ref{figs7} illustrates the evolution of the bipartite entanglement entropy $S_{L/2}$, 
    the probability $f_{\mathrm{skin}}$ and the spin distribution for the circuit with $\delta t=0.5$ 
    and $p=0.7$. The results in Fig.~\ref{figs7}(b) and (c) are evaluated by the above method. 
    We clearly see the existence of the dynamical transition, indicating that the circuit can be used to 
    observe the transition. 
    We have also checked that the results in Fig.~\ref{figs7}(b) and (c) are quantitatively the 
    same as those obtained by averaging the expectation values of these quantities.
    
    To account for the impact of quasidisorder, we incorporate a single-qubit gate on each qubit following 
    the two-qubit unitary operations in each module $M$. 
    The gate on the $l$th qubit realizes the unitary operation $e^{-i\delta t W \mathrm{cos}(2\pi \alpha l) \sigma_l ^z}$, 
    where $W$ and $\alpha$ have the same meaning as defined in Eq. (1) in the main text. 
    For the onsite disorder, one only needs to replace $W \mathrm{cos}(2\pi \alpha l)$ by random numbers 
    drawn from a uniform distribution in the range $[-W,W]$. Figure~\ref{figs8} illustrates that the dynamical 
    transition persists in the presence of weak quasidisorder, implying that our experimental scheme works when quasidisorder is present.
    
    We note that while the quantum circuit may not be exactly mapped to the monitored fermion model discussed in the main text, 
    it can exhibit the same dynamical transition as the fermion model. The quantum circuit can be 
    implemented with superconducting qubits, trapped ions or Rydberg atoms.

\end{widetext}

\end{document}